\documentclass{PoS}
\usepackage{microtype,graphicx,amsmath,amssymb,wrapfig}
\usepackage{fixmath}
\let\OLDthebibliography\thebibliography
\renewcommand\thebibliography[1]{
  \OLDthebibliography{#1}
  \setlength{\parskip}{0pt}
  \setlength{\itemsep}{0pt plus 0.3ex}
}
\title{Theoretical predictions on polarization asymmetry for Drell-Yan process 
       with spin-one deuteron \\ and tensor-polarized structure function $\mathbold{b_1}$}
\ShortTitle{Tensor-polarized structure functions of spin-one deuteron}
\author{S. Kumano$^{\,\, a,b}$ and Qin-Tao Song$^{\, a}$\\
$^a$ KEK Theory Center, Institute of Particle and Nuclear Studies, KEK,\\
\ \ \            and Department of Particle and Nuclear Physics,
Graduate University for Advanced Studies \\
\ \ \  (SOKENDAI),       Ooho 1-1, Tsukuba, Ibaraki, 305-0801, Japan\\ 
$^b$ J-PARC Branch, KEK Theory Center,
     Institute of Particle and Nuclear Studies, KEK,\\
\ \ \ and Theory Group, Particle and Nuclear Physics Division, J-PARC Center,\\
\ \ \ 203-1, Shirakata, Tokai, Ibaraki, 319-1106, Japan\\}

\abstract{
We report recent theoretical progress on 
a polarization asymmetry in the proton-deuteron Drell-Yan process
with a polarized-deuteron target and the tensor-polarized structure 
function $b_1$. Experimental measurements are possible
at JLab for $b_1$ and at Fermilab for the Drell-Yan process.
First, we show a theoretical estimate for the proton-deuteron Drell-Yan
asymmetry in the Fermilab-E1039 experiment. We evolved 
tensor-polarized parton distribution functions,
which explain existing HERMES $b_1$ data,
at $Q^2=2.5$ GeV$^2$ to the $Q^2$ range of the Fermilab Drell-Yan 
measurements. Then, we predicted that the asymmetry is 
of the order of a few percent. The Drell-Yan experiment
has an advantage to probe the tensor-polarized antiquark distributions,
which were suggested by the HERMES experiment as a finite sum for $b_1$
($\int dx b_1 (x) \ne 0$).
Second, we predicted $b_1$ for the JLab experiment by the standard 
convolution model of the deuteron. Our theoretical $b_1$ structure function
seems to be much different from the HERMES data. Furthermore, 
a significant distribution exists at very large $x$ ($>1$) beyond 
the kinematical limit $x_{max}=1$ for the proton. Because the standard 
deuteron-model estimate is much different from the HERMES data, 
there could be an interesting development as a new hadron-physics field 
if future JLab data will be much different from our conventional prediction.
}

\FullConference{23rd international spine symposium (Spin 2018)\\
		September 10-14, 2018\\
		Ferrara, Italy}

\begin{document}

\section{Introduction}
\vspace{-0.2cm}

Origin of nucleon spin has been investigated especially 
from the late 1980's, and major properties became clear. 
On the other hand, spin structure of spin-1 hadrons, such as the deuteron, 
has not been seriously investigated at high energies, although 
electromagnetic properties were studied at low energies such 
as electric quadrupole form factors.
It is possible to probe new spin structure functions 
\cite{fs83,Hoodbhoy:1988am,Kumano:2014pra}
which do not exist in the nucleon. 
There was a HERMES measurement for the deuteron \cite{Airapetian:2005cb};
however, its errors are still large.
There is an approved experiment E12-13-011
at JLab (Thomas Jefferson National Accelerator Facility) 
to measure a tensor-polarized
structure function and it will start soon \cite{Jlab-b1}. 
Furthermore, a polarized proton-deuteron Drell-Yan experiment 
is under consideration in the Fermilab-E1039 experiment \cite{Fermilab-dy}.
Spin physics of spin-1 deuteron could become one of active hadron-physics 
fields to understand it in terms of quark and gluon degrees 
of freedom in the near future.

Due to the spin-one nature, the deuteron has additional 
structure functions in comparison with proton ones, and they
can be investigated in the deep inelastic scattering (DIS) of 
a charged lepton with a polarized deuteron.
The hadron tensor is defined for the spin-1 deuteron as
\cite{Hoodbhoy:1988am,Kumano:2014pra}
\begin{align}
W_{\mu \nu}^{\lambda_f \lambda_i}=&\frac{1}{4 \pi M} \int  d^4x e^{iqx}   
\left \langle p \, \lambda_f |J_{\mu}(x) J_{\nu}(0)   | p \, \lambda_i  \right \rangle
       \notag \\
   = & -F_1 \hat{g}_{\mu \nu} 
     +\frac{F_2}{M \nu} \hat{p}_\mu \hat{p}_\nu 
     + \frac{ig_1}{\nu}\epsilon_{\mu \nu \lambda \sigma} q^\lambda s^\sigma  
     +\frac{i g_2}{M \nu ^2}\epsilon_{\mu \nu \lambda \sigma} 
      q^\lambda (p \cdot q s^\sigma - s \cdot q p^\sigma )
\notag \\
& 
     -b_1 r_{\mu \nu} 
     + \frac{1}{6} b_2 (s_{\mu \nu} +t_{\mu \nu} +u_{\mu \nu}) 
     + \frac{1}{2} b_3 (s_{\mu \nu} -u_{\mu \nu}) 
     + \frac{1}{2} b_4 (s_{\mu \nu} -t_{\mu \nu}) ,
\label{eqn:e1}
\end{align}
where $p$ is the momentum of the deuteron,
$q$ is the momentum of the virtual photon,  and
$\lambda_i$ and $\lambda_f$ indicate initial and final spin states
of the deuteron, respectively.
The notations
$\hat{g}_{\mu \nu}$ and $\hat{p}_\mu$ are defined as
$\hat{g}_{\mu \nu} \equiv  g_{\mu \nu} -{q_\mu q_\nu}/{q^2}$ and 
$\hat{p}_\mu \equiv p_\mu - ({p \cdot q}/{q^2}) \, q_\mu$,
$s^\mu$ is the spin vector of the deuteron, and one can find
definitions of other kinematical variables
in Refs. \cite{Hoodbhoy:1988am,Kumano:2014pra}.  
In Eq.\,(\ref{eqn:e1}), there are 8 structure functions in total. 
The structure functions $F_1$, $F_2$, $g_1$ and $g_2$ appear  
in the hadron tensor of the proton, whereas the structure functions 
$b_1$, $b_2$, $b_3$ and $b_4$ are the new ones for the deuteron.
The leading-twist structure functions are $b_1$ and $b_2$, and 
they are related with each other by the Callan-Gross like relation 
$b_2 = 2 x b_1$. The functions $b_3$ and $b_4$ are higher-twist ones. 

As the first step, we may investigate the leading-twist function $b_1$
among the tensor-polarized structure functions $b_{1-4}$.
The function $b_1$ will clarify the tensor structure of the deuteron
in terms of quarks and gluons, 
and it is expressed by the tensor-polarized parton distribution functions
(PDFs) $\delta_Tq_i $ as
\begin{align}
b_1 (x) = \frac{1}{2 }\sum_i e_i^2 
 \left [ \delta_Tq_i (x)+\delta_T \bar q_i(x) \right ] , \ \
\delta_Tq_i (x) \equiv q^0_i (x) -\frac{q_i^{+1}(x) +q_i^{-1} (x)}{2} ,
\label{eqn:e7}
\end{align}
where the index $i$ is the quark flavor, the superscripts
(0, $\pm$) indicate the deuteron spin state,
and the scale-$Q^2$ dependence is abbreviated for simplicity.
There is an interesting sum rule of  $b_1$ \cite{Close:1990zw}:
\begin{align}
\int dx \, b_1(x) = 
- \lim_{t \to 0} \, \frac{5}{24} \, t \, F_Q (t) 
+ \frac{1}{9} \! \int \! dx 
\left[  4  \delta_T \bar u(x)+  4  \delta_T \bar d(x)  
+  \delta_T \bar s(x)   \right ] ,
\label{eqn:e8}
\end{align}
where $F_Q (t)$ is the electric quadrupole form factor. 
Because the first term vanishes, the nonzero integral 
of $b_1$ indicates the existence of finite tensor-polarized 
antiquark distributions.

The first measurement of $b_1$ was conducted by 
the HERMES collaboration \cite{Airapetian:2005cb}, 
and the values of $b_1$ are of the order of $10^{-2}$.
However, theoretical predictions of $b_1$ are much smaller 
than the experimental measurements \cite{Khan:1991qk,Cosyn:2017fbo}
as we show later in Sec.\,\ref{b1-results} by
conventional convolution models.
The HERMES collaboration also reported the integrals of $b_1$ as
\begin{align}
\! \! \!
\int _{0.002}^{0.85} \! dx \, b_1(x)&=
 \left[0.35\pm 0.10(\text{stat}) \pm 0.18(\text{sys})  \right ]\times10^{-2} 
      \ \ \text{(in the range of $Q^2>1$ GeV$^2$)} ,
\label{eqn:e3}
\end{align}
and it was $ [1.05\pm 0.34(\text{stat})  \pm 0.35(\text{sys}) ]\times10^{-2} $
in the whole measured range.
These HERMES measurements seem to indicate the existence 
of finite tensor-polarized antiquark distributions
according to Eq.\,(\ref{eqn:e8}). 
However, such tensor polarization in antiquarks may not be
easily understood theoretically in simple deuteron models.
In the near future, $b_1$ will be measured by the experiment E12-13-011 
at JLab, and it could clarify the tensor structure in terms of 
quark and gluon degrees of freedom. However, 
the antiquark distributions can be measured more directly
by the Drell-Yan process as discussed in the next section.

\section{Theoretical estimate on tensor-polarization asymmetry 
in proton-deuteron Drell-Yan process}

\begin{wrapfigure}[10]{r}{0.43\textwidth}
   \vspace{-0.4cm}
   \hspace{0.15cm}
     \includegraphics[width=6.0cm]{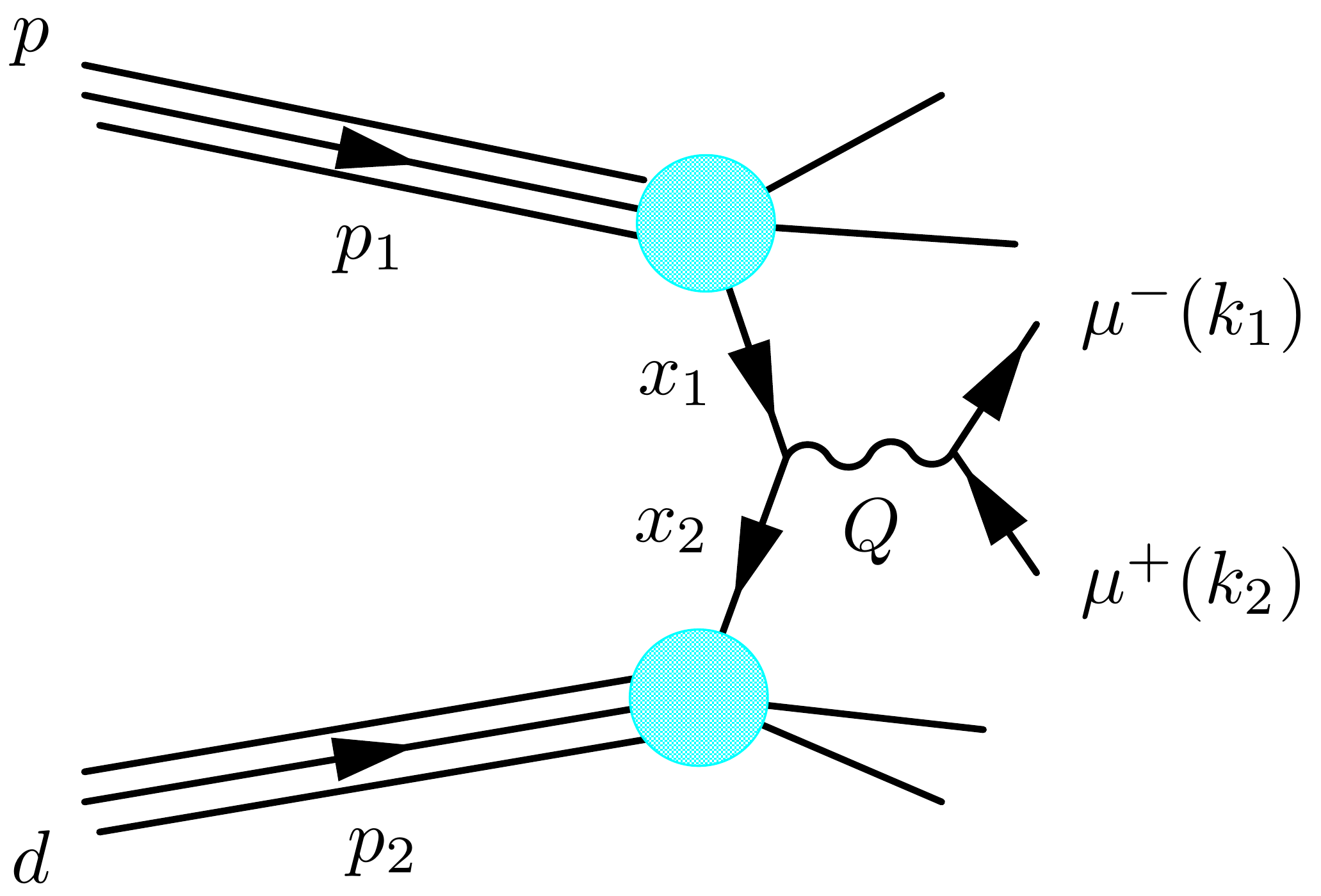}
\vspace{-0.3cm}
    \caption{\label{fig:pd} Proton-deuteron Drell-Yan process.}
\label{fig:dy}
\vspace{-0.5cm}
\end{wrapfigure}

\vspace{-0.2cm}
The tensor structure of the deuteron can be investigated
by the proton-deuteron Drell-Yan process, which is possible 
at Fermilab. There is a significant advantage in probing
the tensor-polarized antiquark distributions 
by the Drell-Yan process illustrated in Fig.\,\ref{fig:dy}.
At Fermilab, the beam is unpolarized 120 GeV proton provided 
by the Main Injector and the deuteron target is tensor polarized. 
The momentum fractions carried by the quark and antiquark are 
denoted as $x_1$ and $x_2$, and the scale is given by
$Q^2=x_1 x_2 s$ with the center-of-mass-energy squared $s=(p_1+p_2)^2$.

The hadron tensor of the Drell-Yan process is complicated 
than that of DIS since there are more structure functions involved
\cite{Hino-Kumano}. 
Among spin asymmetries, the tensor-polarization asymmetry $A_Q$ provides 
information of the tensor-polarized PDFs, and it is defined as 
\begin{align}
A_{Q}=\frac{1}{\left \langle \sigma  \right \rangle} 
\left [ \sigma(\bullet, 0)-\frac{\sigma(\bullet, +1)
        +\sigma(\bullet, -1)}{2}  \right],
\label{eqn:e4}
\end{align}
where $\bullet$ represents unpolarized proton beam and the deuteron spin states
are denoted as $\pm$ and $0$. The spin asymmetry $A_Q$ indicates the difference 
of the cross section with different deuteron spin states. 
In the parton model, $A_Q$ can be expressed by the tensor-polarized PDFs as
\cite{Hino-Kumano}
\begin{align}
A_{Q}=\frac{\sum_i e_i^2 \left [q_i(x_1) \delta_T \bar q_i(x_2)
+\bar q_i(x_1) \delta_T q_i(x_2) \right]}{\sum_i e_i^2 
\left [q_i(x_1)\bar q_i(x_2)+\bar q_i(x_1)  q_i(x_2) \right]} ,
\label{eqn:e4}
\end{align}
where the scale $Q^2$ is abbreviated.
If  $x_F=x_1-x_2$ is large enough, the contribution of 
$\bar q_i(x_1) \delta_T q_i(x_2)$ can be neglected in comparison with
$q_i(x_1) \delta_T \bar q_i(x_2)$, so it is possible to probe
the tensor-polarized antiquark distributions
$\delta_T \bar q_i (x)$ by using the Drell-Yan process.

In order to predict the spin asymmetry in the Drell-Yan process at Fermilab,
the tensor-polarized distributions $\delta_T q_i(x, Q^2)$ are needed.
Here, we adopt the parameterizations of  $\delta_T q_i(x, Q^2_0)$ in 
Ref. \cite{Kumano:2010vz}, where there are two sets of 
tensor-polarized distributions $\delta_T q_i(x, Q^2_0)$ based 
on the analysis of HERMES data at the average scale $Q^2_0=2.5$ GeV$^2$. 
In the set-1 analysis, there are no tensor-polarized antiquark distributions 
at the initial scale $Q^2_0=2.5$ GeV$^2$; however, 
finite tensor-polarized antiquark distributions are allowed
in the set-2 analysis. With $\delta_T q_i(x, Q^2_0)$ at the initial scale, 
one can obtain the tensor-polarized distributions at lager $Q^2$ 
by using DGLAP (Dokshitzer-Gribov-Lipatov-Altarelli-Parisi) 
evolution equations \cite{Hoodbhoy:1988am,MK-1996}.

\begin{wrapfigure}[11]{r}{0.41\textwidth}
   \vspace{-0.2cm}
     \includegraphics[width=6.0cm]{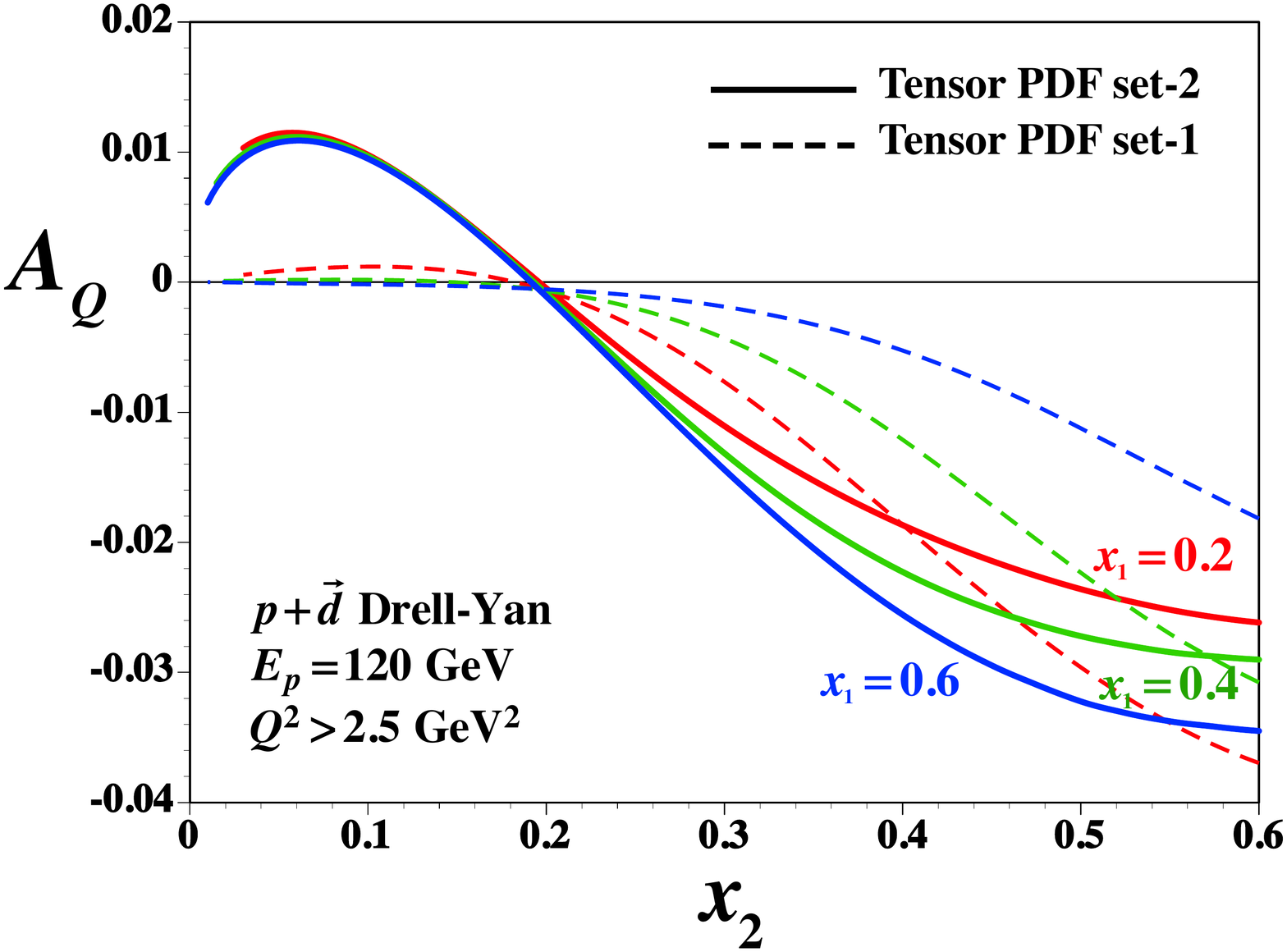}
\vspace{-0.4cm}
    \caption{\label{fig:dyaq} Spin asymmetries $A_{Q}$  
    at $x_1=0.2$, $x_1=0.4$ and $x_1=0.6$ by set 1 and 2
     \cite{Kumano:2016ude}.}
\label{fig:dyaq}
\vspace{-0.4cm}
\end{wrapfigure}

In Fig.\,\ref{fig:dyaq}, the spin asymmetries $A_{Q}$ are shown 
for the Drell-Yan experiment at Fermilab,
and the momentum fraction $x_1$ is fixed as $x_1=0.2$, $x_1=0.4$ 
and $x_1=0.6$ in both set 1 and set 2 \cite{Kumano:2016ude}. 
The spin asymmetries are not large, and they are typically 
of order of a few percent. The spin asymmetries of set 1 and set 2 
are very different at the small $x_2$, 
since the term $q_i(x_1) \delta_T \bar q_i(x_2)$
is dominant in this region (large $x_F=x_1-x_2$) and 
$ \delta_T \bar q_i(x, Q^2_0) =0$ in the set 1. 
Because the set-2 analysis provides a better description 
of the HERMES measurements, the spin asymmetries of the set 2 
should be more reliable than those of the set 1. In future, 
the spin asymmetry $A_{Q}$ could be measured by 
the Fermilab-E1039 (SpinQuest) collaboration in the Drell-Yan process 
with tensor-polarized deuteron target at Fermilab, 
and our theoretical predictions provide baseline for planning
the experimental measurement.

\vspace{-0.30cm}
\section{Standard convolution model prediction for $\mathbold{b_1}$}
\label{convolution}
\vspace{-0.20cm}

A convolution formalism has been used for describing nuclear structure
functions at medium and large $x$ ($x>0.2$) as a standard description
to explain nuclear modifications in terms of nuclear binding 
and nucleon Fermi motion. A nuclear structure function is expressed
by the nucleonic structure function convoluted with 
a spectral function which indicates a nucleon momentum 
distribution in a nucleus. We use this description to 
calculate $b_1$ for the deuteron. 
Specifically, we employ two convolution models \cite{Cosyn:2017fbo}
for calculating $b_1$ with D-state admixture in the deuteron.
One is a basic convolution description,
and the other is a virtual nucleon approximation 
which includes higher-twist contributions.

\vspace{-0.30cm}
\subsection{Basic convolution description (Theory 1)}
\label{convolution}
\vspace{-0.20cm}

In a basic convolution model for nuclear structure functions,
the nuclear tensor $W_{\mu\nu}^A$ is given by the nucleonic one $W_{\mu\nu}$
convoluted with the nucleon's momentum distribution expressed by
the spectral function $S (p)$ as
\vspace{-0.20cm}
\begin{align}
& \! \! \! \!
W_{\mu\nu}^A (P_A, q)  = \int d^4 p \, S(p) \, W_{\mu\nu} (p, q) ,
\ \ \ 
S (p) = \frac{1}{A} \sum_i
      | \phi_i (\vec p \,) |^2 
     \delta \bigg ( p^0-M_A + \sqrt{M_{A-i}^{\ 2} 
                    + \vec p^{\ 2}} \, \bigg ) .
\label{eqn:w-convolution}
\\[-0.90cm] \nonumber
\end{align}
Here, $p$ and $P_A$ are momenta for the nucleon and nucleus,
and $\phi_i (\vec p \,)$ is the momentum-space wave function
with the nucleon index $i$.
This description has been successful in explaining gross features
of nuclear modifications at medium and large $x$ ($x>0.2$).
Physics mechanisms are contained in the spectral function
as nuclear binding, Fermi motion, and short-range correlations.

For extracting the structure function $b_1$ from the deuteron tensor 
$W_{\mu\nu}^D (p_{_D},q)$, helicity amplitudes are defined 
by the photon polarization vector $\varepsilon_h^{\mu}$ as
$ A_{hH,hH} (x,Q^2)  
   =  \varepsilon_h^{*\mu} \varepsilon_h^{\nu} \, 
       W_{\mu\nu}^D (p_{_D},q) $
\cite{Hoodbhoy:1988am}
and in the same way as $\hat A_{hs,hs} (x,Q^2)$ for the nucleon.
In the Bjorken scaling limit, the structure function $b_1$ 
of the deuteron and $F_1$ of the nucleon 
are expressed by the helicity amplitudes as
\cite{Hoodbhoy:1988am,kk08}
\vspace{-0.10cm}
\begin{align}
b_1  = A_{+0,+0} - \frac{A_{++,++} +A_{+-,+-}}{2} , \ \ \ 
F_1^N = \frac{ A_{+\uparrow,+\uparrow} +A_{+\downarrow,+\downarrow} }{2} .
\label{eqn:A-b1F1}
\\[-0.90cm] \nonumber
\end{align}
From these relations, the deuteron $b_1$ is expressed by the convolution
integral with the unpolarized structure function $F_1^N$ for the nucleon as
\vspace{-0.10cm}
\begin{align}
b_1 (x,Q^2)  = \int \frac{dy}{y} 
\, \delta_T f(y) \, F_1^N (x/y,Q^2), 
\ \ \ 
\delta_T f(y)  \equiv f^0 (y)  - \frac{f^+ (y) + f^- (y)}{2} ,
\label{eqn:b1-convolution}
\\[-0.90cm] \nonumber
\end{align}
where $b_1$ is defined by the one per nucleon.
The polarized structure function $b_1$ is given by
``unpolarized" parton distributions in the tensor-polarized deuteron
as defined in Eq.\,(\ref{eqn:e7}), so that $b_1$ is expressed 
by the unpolarized $F_1^N$ in the convolution formalism.
The function $f^H (y)$ is the lightcone momentum distribution 
with the deuteron spin state $H$, and it is expressed 
by the deuteron wave function $\phi^H (\vec p \,)$ as
$ f^H (y) = \int d^3 p \, y \, | \, \phi^H (\vec p \,) \, |^2
          \, \delta \left ( y - \frac{E-p_z}{M_N}   \right ) $.
The momentum fraction $y$ is defined by
$ y   =  M \, p \cdot q / (M_{N} \, P \cdot q ) \simeq 2 \, p^- / P^- $
where the light-cone coordinate $p^-$ is given by
$\, p^- \equiv (p^0 -p^3)/\sqrt{2} \,$ 
with the $z$-axis along the virtual-photon momentum direction.
Expressing S- and D-state wave functions as $\phi_0 (p)$ and $\phi_2 (p)$,
respectively, we finally obtain the lightcone-momentum distribution as
\vspace{-0.00cm}
\begin{align}
\delta_T f(y)  = \int d^3 p \, y 
 \left [ - \frac{3}{4 \sqrt{2} \pi} \phi_0 (p) \phi_2 (p) 
  + \frac{3}{16\pi} |\phi_2 (p)|^2 \right ]
 (3 \cos^2 \theta -1) \, \delta \left ( y - \frac{p\cdot q}{M_N \nu}  \right ) .
\vspace{-0.00cm}
\label{eqn:delta-t-f}
\end{align}
It is clear in this expression that $b_1 (x)$ vanishes if 
there is no D-wave admixture ($\phi_2 =0$) in the convolution description. 
The first term of Eq.\,(\ref{eqn:delta-t-f}) comes from the S-D interference
and the second one does purely from the D state.

\vspace{-0.30cm}
\subsection{Virtual nucleon approximation (Theory 2)}
\label{vna-model}
\vspace{-0.10cm}

Next, we calculate $b_1$ in a virtual nucleon approximation
by including higher-twist effects, whereas
the scaling-limit relations of Eq.\,(\ref{eqn:A-b1F1})
are used in the first model to obtain
Eq.\,(\ref{eqn:delta-t-f}).
The DIS cross section of charged-lepton with the polarized deuteron
is written as
\vspace{-0.10cm}
\begin{align}
\frac{d\sigma}{dx \, dQ^2} 
= & \frac{\pi y^2\alpha^2}{Q^4(1-\epsilon)}
\bigg[ F_{UU,T}+\epsilon F_{UU,L}
+T_{\parallel\parallel} \left( F_{UT_{LL},T}+\epsilon F_{UT_{LL},L} \right) 
\nonumber  \\[-0.20cm]
& \ \ \ \ \ \ \ \ \ \ \ \ \ \ \ \ 
+ T_{\parallel\perp} \cos\phi_{T_\parallel} \sqrt{2\epsilon(1+\epsilon)} \,
   F_{UT_{LT}}^{\cos\phi_{T_\parallel}} 
+ T_{\perp\perp} \cos(2\phi_{T_\perp}) \, \epsilon \,
   F_{UT_{TT}}^{\cos(2\phi_{T_\perp})}\bigg ]\, .
\label{eq:cross}
\\[-0.80cm] \nonumber
\end{align}
Here, $\epsilon$ is the degree of the longitudinal polarization 
of the virtual photon.
The details of the polarization factors
($T_{\parallel\parallel}$, $T_{\parallel\perp}$, $T_{\perp\perp}$)
and the angles ($\phi_{T_\parallel}$, $\phi_{T_\perp}$)
are explained in Ref.\,\cite{CSW}.
Among these structure functions, $b_1$ is related to 
$ F_{UT_{LL},T} $, $ F_{UT_{TT}}^{\cos(2\phi_{T_\perp})} $, 
and the helicity amplitudes as
\vspace{-0.20cm}
\begin{align}
& b_1  = - \frac{1}{1+\gamma^2} \sqrt{\frac{3}{8}} \,
   \left [ F_{UT_{LL},T}
    +F_{U\mathcal{T}_{TT}}^ { \cos(2\phi_ { T_\perp }) } \right ] ,
\nonumber \\[-0.20cm]
& \ \ \ \ \ \ 
F_{UT_{LL},T} 
     =\frac{2}{\sqrt{6}} \left(A_{++,++}-2A_{+0,+0}+A_{+-,+-}\right) , \ \ \ 
F_{UT_{TT}}^{\cos(2\phi_{T_\perp})} = -\sqrt{\frac{2}{3}} \Re e A_{+-,-+} \ ,
\label{eq:b1_f}
\\[-0.90cm] \nonumber
\end{align}
where the factor $\gamma$ is defined by $\gamma =\sqrt{Q^2}/\nu$.

Now, we use the virtual nucleon approximation (VNA) for calculating
the deuteron tensor and subsequently $b_1$.
Let us consider the $np$ component of the light-front deuteron wave function.
In this model, the virtual photon interacts with an off-shell nucleon
and another non-interacting spectator nucleon is assumed to be on mass shell.
Then, the deuteron tensor is calculated by integrating over 
the spectator momentum $\vec p_N$:
\vspace{-0.20cm}
\begin{align}
& W_{\mu\nu}^{\lambda' \lambda} (P,q)
 \! = \! 4(2\pi)^3 \! \! \int \! d\Gamma_N
\frac{\alpha_{_N}}{\alpha_i} 
W_{\mu\nu} (p_{i},q)
\rho_D(\lambda',\lambda) ,
\label{eq:w-convolution}
\\[-0.90cm] \nonumber
\end{align}
where $d\Gamma_N$ is the phase space for the spectator nucleon.
The variables $ \alpha_i $ and $ \alpha_{_N} $ are the momentum fractions 
for the interacting ($i$) and spectator ($N$) nucleons
defined by
$ \alpha_i= 2 \, p_i^- / P^-$ and
$ \alpha_{_N}= 2 \, p_N^- / P^-=2-\alpha_i $.
The deuteron density $\rho_D(\lambda',\lambda)$ is given by
the deuteron wave function $\Psi^{D}_\lambda(\vec{k},\lambda'_N,\lambda_N)$
expressed by the S- and D-state wave components $\phi_0$ and $\phi_2$ 
\cite{Cosyn:2017fbo}.
The ratio $\alpha_{_N}/\alpha_i$ comes from the fact that
the hadron tensor $W_{\mu\nu}$ is for the nucleon with momentum $p_i$
rather than the one at rest.
Calculating the relations in Eq.\,(\ref{eq:b1_f}), we obtain
$b_1$ in the VNA model as
\vspace{-0.20cm}
\begin{align}
b_1(x, Q^2) = \frac{3}{4(1+\gamma^2)} & \int 
                 dk \, d(\cos\theta_k) \, \frac{k^2}{\alpha_i}
\bigg[ F_{1}^N(x_i,Q^2) \left(6\cos^2\theta_k-2\right) 
\nonumber \\[-0.10cm]
& \ \ \ \ \ 
-\frac{T^2 } {2 \, p_i \cdot q } \, F_ { 2 }^N (x_i ,Q^2)
\left(5\cos^2\theta_k-1\right) \bigg]
\left[ - \frac{\phi_0 (k) \phi_2(k)}{\sqrt{2}}+\frac{\phi_2( k)^2}{4}\right] ,
\label{eq:b1_vna}
\\[-0.80cm] \nonumber
\end{align}
where $T^\mu$ is defined by
$ T^\mu  = p_N^\mu+ q^\mu p_N\cdot q / Q^2 - L^\mu p_N\cdot L / L^2$ with
$ L^\mu=P^\mu+ q^\mu P\cdot q / Q^2$.

\vspace{-0.30cm}
\subsection{Results on $\mathbold{b_1}$}
\label{b1-results}
\vspace{-0.10cm}

\begin{figure}[b!]
\vspace{-0.05cm}
\begin{minipage}{\textwidth}
\begin{tabular}{lc}
\hspace{-0.30cm}
\begin{minipage}[c]{0.45\textwidth}
   \vspace{-0.2cm}
   \begin{center}
    \includegraphics[width=5.5cm]{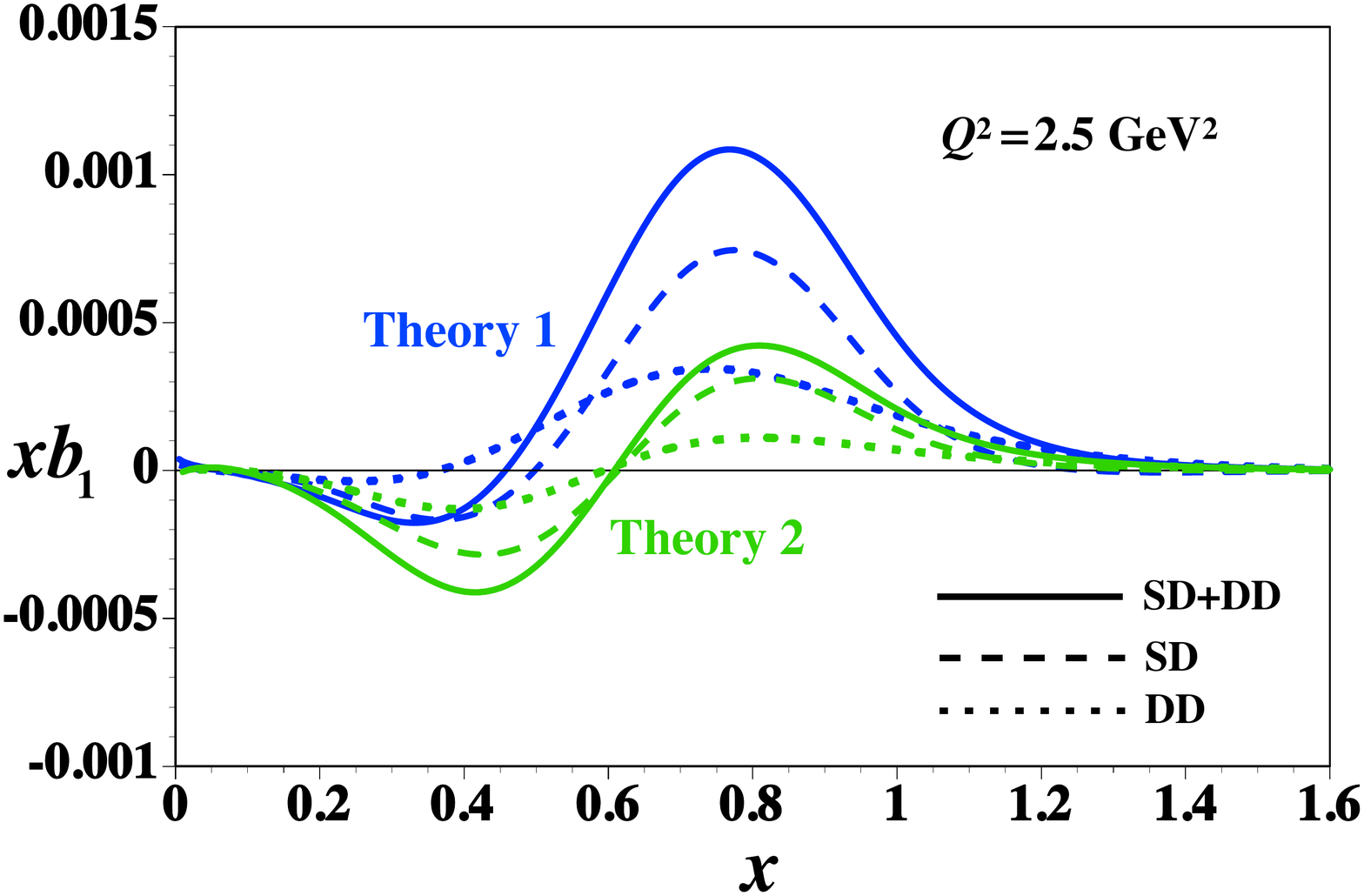}
   \end{center}
\vspace{-0.75cm}
\caption{Convolution model for $b_1$ \cite{Cosyn:2017fbo}.}
\label{fig:b1-convolution}
\vspace{-0.4cm}
\end{minipage} 
\hspace{0.5cm}
\begin{minipage}[c]{0.45\textwidth}
    \vspace{-0.1cm}
   \begin{center}
    \includegraphics[width=5.5cm]{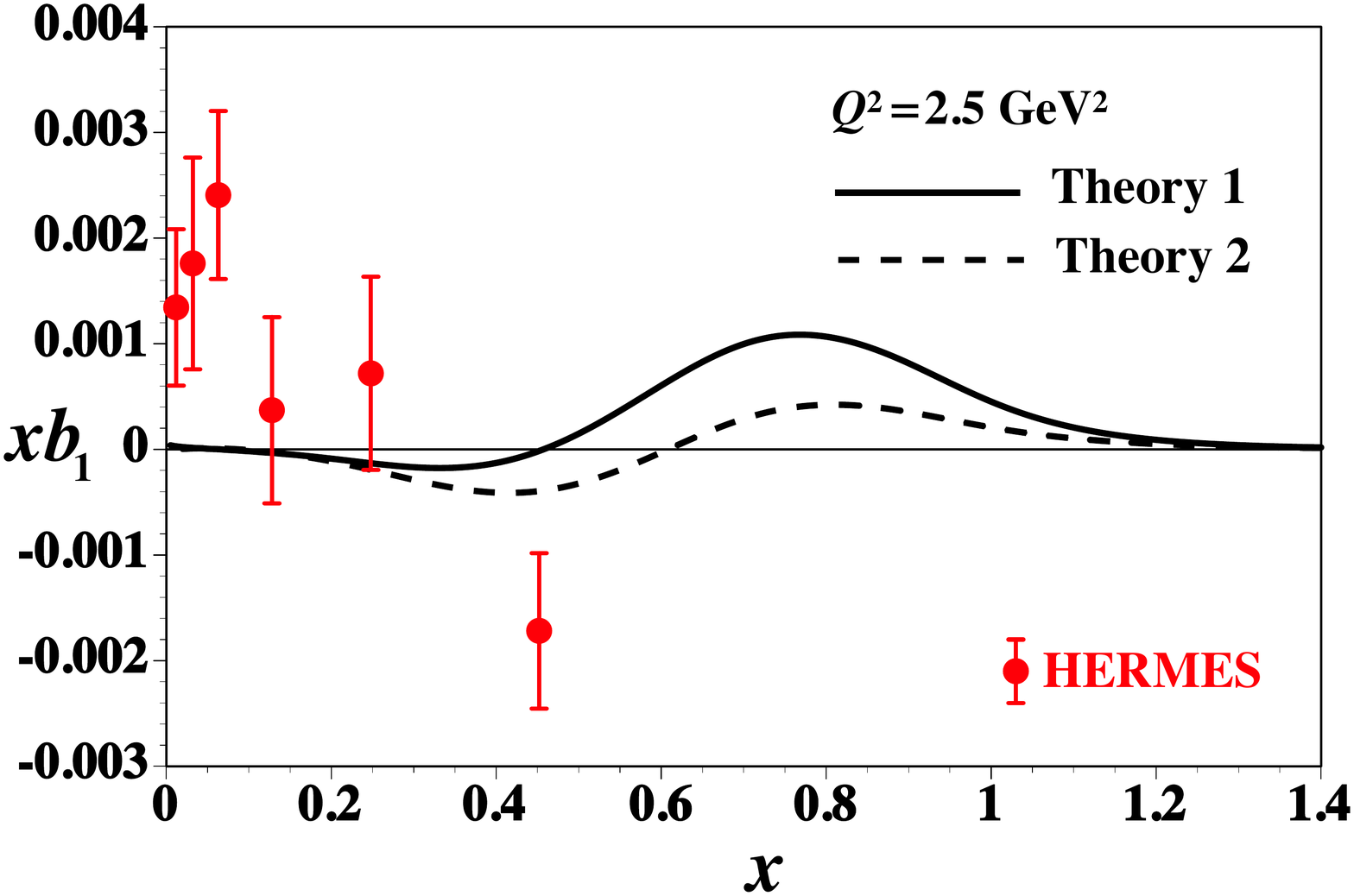}
   \end{center}
\vspace{-0.80cm}
\caption{Comparison with HERMES data \cite{Cosyn:2017fbo}.}
\label{fig:b1-hermes-convolution}
\vspace{-0.4cm}   
\end{minipage}
\end{tabular}
\vspace{0.20cm}
\end{minipage}
\end{figure}

We show numerical results for $b_1$ by using Eqs.\,(\ref{eqn:b1-convolution})
and (\ref{eq:b1_vna}) in Fig.\,\ref{fig:b1-convolution}
at $Q^2=2.5$ GeV$^2$, which is the average scale of the HERMES measurement
\cite{Airapetian:2005cb}.
As for the nucleon structure functions, we used
the MSTW2008 (Martin-Stirling-Thorne-Watt, 2008) 
leading-order (LO) parton distributions and 
the SLAC-R1998 parametrization for the longitudinal-transverse ratio $R$.
The CD-Bonn model was employed for the deuteron wave function.
The S-D interference contributions, D-wave ones, 
and their total distributions are shown. The S-D terms are larger than 
the D terms in both theoretical calculations of 1 and 2. However, 
the D terms are larger, in comparison with the S term,
than expected from the D-wave admixture probability of several percent.
There are some differences between two theory results. 
They come from mainly higher-twist effects, but there are also
effects coming from slightly different normalizations 
in the lightcone wave functions.
Both results are also very different from previous convolution 
calculations in Refs.\,\cite{Hoodbhoy:1988am,Khan:1991qk} although 
the theoretical formalisms are similar.
First, the $x$ dependence is very different. Especially,
the SD terms have opposite sign to the one in Ref.\,\cite{Khan:1991qk}.
The large-$x$ distributions exist even at $x>1$, whereas
there is no distribution in Ref.\,\cite{Khan:1991qk}.

The total $b_1$ distributions are compared with the HERMES data
in Fig.\,\ref{fig:b1-hermes-convolution}.
In comparison with the HERMES $xb_1 \sim (1\text{-}3 \times 10^{-3})$,
the theoretical distributions are rather small and 
much less than $10^{-3}$ at $x<0.6$. 
Due to the large experimental errors, we cannot conclude
whether significant differences actually exist between the data 
and conventional theoretical estimates at this stage.
However, the large differences may indicate a new hadron physics 
mechanism for explaining the experimental measurements \cite{miller-b1}, 
although the differences may also come from higher-twist effects.
It is interesting to find the large differences between
the HERMES data and the standard convolution calculations.
In the near future, the JLab experiment will start to
measure accurately $b_1$ at medium $x$ ($0.3<x<0.5$) \cite{Jlab-b1},
and there is a possibility to measure the proton-deuteron Drell-Yan
process in the Fermilab-E1039 experiment 
\cite{Fermilab-dy,Kumano:2016ude}.
The tensor structure functions are interesting topics
in 2020's for probing a new aspect of high-energy hadron physics.

\vspace{-0.3cm}
\section{Summary}
\vspace{-0.30cm}

We estimated the tensor-polarization asymmetry in the proton-deuteron
Drell-Yan process for a possible Fermilab-E1039 experiment. Using
the tensor-polarized PDFs for explaining the HERMES data, we obtained
that the asymmetry is of the order of a few percent. Because a finite antiquark
tensor polarization was suggested in the HERMES experiment by using
the $b_1$ sum rule, this Drell-Yan experiment is an interesting one
to shed light on a new aspect of hadron physics as the tensor-polarized
antiquark distributions. Next, we showed standard deuteron calculations
on $b_1$ by using the convolution descriptions. We found that 
the theoretical $b_1$ distributions are much different from 
the HERMES data and that a significant $b_1$ distribution
exists at large $x$ (even $x>1$). Since a new $b_1$ experiment will
start soon at JLab, the tensor-polarized structure functions will be
interesting hadron-physics topics in 2020's.

\vspace{-0.30cm}
\acknowledgments
\vspace{-0.30cm}
\noindent 
Q.-T. S is supported by the MEXT Scholarship for foreign students 
through the Graduate University for Advanced Studies.

\vspace{-0.25cm}



\begin{thebibliography}{99}
\vspace{-0.30cm}
\bibitem{fs83}  L. L. Frankfurt and M. I. Strikman, 
                   Nucl. Phys. A {\bf 405} (1983) 557.
\bibitem{Hoodbhoy:1988am}
  P.~Hoodbhoy, R.~L.~Jaffe and A.~Manohar,
  Nucl.\ Phys.\ B {\bf 312} (1989) 571.
\bibitem{Kumano:2014pra}
  S.~Kumano, J.\ Phys.\ Conf.\ Ser.\  {\bf 543} (2014) 012001.
\bibitem{Airapetian:2005cb}
  A.~Airapetian {\it et al.} [HERMES Collaboration],
  Phys.\ Rev.\ Lett.\  {\bf 95} (2005) 242001.
\bibitem{Jlab-b1} JLab-E12-13-011 experiment, 
                  Jefferson Lab PAC-40,
                  K. Allada {\it et al.} (2013).
\bibitem{Fermilab-dy} 
 The polarized proton-deuteron Drell-Yan measurement is considered
 in the Fermilab E1039 experiment,
 Letter of Intent Report No. P1039 (2013).
\bibitem{Close:1990zw}
  F.~E.~Close and S.~Kumano,
  Phys.\ Rev.\ D {\bf 42} (1990) 2377.  
\bibitem{Khan:1991qk}
  H.~Khan and P.~Hoodbhoy,
  Phys.\ Rev.\ C {\bf 44} (1991) 1219.
\bibitem{Cosyn:2017fbo} 
  W.~Cosyn, Y.~B.~Dong, S.~Kumano and M.~Sargsian,
  Phys.\ Rev.\ D {\bf 95} (2017) 074036.
\bibitem{Hino-Kumano}
  S.~Hino and S.~Kumano,
  Phys.\ Rev.\ D {\bf 59} (1999) 094026;
  {\bf 60} (1999) 054018.
\bibitem{Kumano:2010vz}
  S.~Kumano, Phys.\ Rev.\ D {\bf 82} (2010) 017501.
\bibitem{MK-1996}
  M. Miyama and S. Kumano, Comput. Phys. Commun. {\bf 94} (1996) 185.
\bibitem{Kumano:2016ude}
  S.~Kumano and Q.~T.~Song,
  Phys.\ Rev.\ D {\bf 94} (2016) 054022.
\bibitem{kk08} 
  T.-Y. Kimura and S. Kumano, 
  Phys. Rev. D {\bf 78} (2008) 117505.
\bibitem{CSW} 
  W. Cosyn, M. Sargsian, and C. Weiss, to be submitted for publication.
\bibitem{miller-b1} 
      G. A. Miller,  Phys. Rev. C {\bf 89} (2014) 045203.  
\end{thebibliography}
\end{document}